\input lanlmac.tex

%\def\subsubsec#1{$\underline{\rm #1}$}

% Something to deal with sub-sub-sections

\def\unlockat{\catcode`\@=11}
\def\lockat{\catcode`\@=12}

\unlockat
% Something to deal with sub-sub-sections

\def\newsec#1{\global\advance\secno by1\message{(\the\secno. #1)}
\global\subsecno=0\global\subsubsecno=0\eqnres@t\noindent
{\bf\the\secno. #1}
\writetoca{{\secsym} {#1}}\par\nobreak\medskip\nobreak}
\global\newcount\subsecno \global\subsecno=0
\def\subsec#1{\global\advance\subsecno
by1\message{(\secsym\the\subsecno. #1)}
\ifnum\lastpenalty>9000\else\bigbreak\fi\global\subsubsecno=0
\noindent{\it\secsym\the\subsecno. #1}
\writetoca{\string\quad {\secsym\the\subsecno.} {#1}}
\par\nobreak\medskip\nobreak}
\global\newcount\subsubsecno \global\subsubsecno=0
\def\subsubsec#1{\global\advance\subsubsecno by1
\message{(\secsym\the\subsecno.\the\subsubsecno. #1)}
\ifnum\lastpenalty>9000\else\bigbreak\fi
\noindent\quad{\secsym\the\subsecno.\the\subsubsecno.}{#1}
\writetoca{\string\qquad{\secsym\the\subsecno.\the\subsubsecno.}{#1}}
\par\nobreak\medskip\nobreak}

\def\subsubseclab#1{\DefWarn#1\xdef
#1{\noexpand\hyperref{}{subsubsection}%
{\secsym\the\subsecno.\the\subsubsecno}%
{\secsym\the\subsecno.\the\subsubsecno}}%
\writedef{#1\leftbracket#1}\wrlabeL{#1=#1}}% Macros for boxes
\lockat

\def\IL{\relax{\rm I\kern-.18em L}}
\def\IH{\relax{\rm I\kern-.18em H}}
\def\IR{\relax{\rm I\kern-.18em R}}
\def\IC{\relax\hbox{$\inbar\kern-.3em{\rm C}$}}
\def\IZ{\relax\ifmmode\mathchoice
{\hbox{\cmss Z\kern-.4em Z}}{\hbox{\cmss Z\kern-.4em Z}}
{\lower.9pt\hbox{\cmsss Z\kern-.4em Z}}
{\lower1.2pt\hbox{\cmsss Z\kern-.4em Z}}\else{\cmss Z\kern-.4em
Z}\fi}
\def\CM {{\cal M}}

\def\CB {{\cal B}}

%% MORE MACROS
\def\CM {{\cal M}}

\font\manual=manfnt \def\dbend{\lower3.5pt\hbox{\manual\char127}}

\def\IZ{\relax\ifmmode\mathchoice
{\hbox{\cmss Z\kern-.4em Z}}{\hbox{\cmss Z\kern-.4em Z}}
{\lower.9pt\hbox{\cmsss Z\kern-.4em Z}}
{\lower1.2pt\hbox{\cmsss Z\kern-.4em Z}}\else{\cmss Z\kern-.4em
Z}\fi}

\def\CM {{\cal M}}

% more macros, alphabetically

\def\IZ{\relax\ifmmode\mathchoice
{\hbox{\cmss Z\kern-.4em Z}}{\hbox{\cmss Z\kern-.4em Z}}
{\lower.9pt\hbox{\cmsss Z\kern-.4em Z}}
{\lower1.2pt\hbox{\cmsss Z\kern-.4em Z}}\else{\cmss Z\kern-.4em
Z}\fi}
\def\IB{\relax{\rm I\kern-.18em B}}
\def\IC{{\relax\hbox{$\inbar\kern-.3em{\rm C}$}}}
\def\ID{\relax{\rm I\kern-.18em D}}
\def\IE{\relax{\rm I\kern-.18em E}}
\def\IF{\relax{\rm I\kern-.18em F}}
\def\IG{\relax\hbox{$\inbar\kern-.3em{\rm G}$}}
\def\IGa{\relax\hbox{${\rm I}\kern-.18em\Gamma$}}
\def\IH{\relax{\rm I\kern-.18em H}}
\def\II{\relax{\rm I\kern-.18em I}}
\def\IK{\relax{\rm I\kern-.18em K}}
\def\IP{\relax{\rm I\kern-.18em P}}

\def\inbar{\,\vrule height1.5ex width.4pt depth0pt}

\font\cmss=cmss10 \font\cmsss=cmss10 at 7pt
\def\IR{\relax{\rm I\kern-.18em R}}

% Macros for boxes

\def\boxit#1{\vbox{\hrule\hbox{\vrule\kern8pt
\vbox{\hbox{\kern8pt}\hbox{\vbox{#1}}\hbox{\kern8pt}}
\kern8pt\vrule}\hrule}}
\def\mathboxit#1{\vbox{\hrule\hbox{\vrule\kern8pt\vbox{\kern8pt
\hbox{$\displaystyle #1$}\kern8pt}\kern8pt\vrule}\hrule}}

%% ANOTHER SET OF MACROS

\def\inbar{\,\vrule height1.5ex width.4pt depth0pt}

\font\cmss=cmss10 \font\cmsss=cmss10 at 7pt
\def\IR{\relax{\rm I\kern-.18em R}}

\Title{ \vbox{\baselineskip12pt\hbox{hep-th/9610150}
\hbox{YCTP-P22-96} \hbox{TMUP-HEL-9611}}}
{\vbox{
\centerline{The Thermodynamic Bethe Ansatz and the}
\centerline{1/N Correction to the Density Phase Transition}
\centerline{in the Gross-Neveu Model}
}}\footnote{}
\medskip
\centerline{Alan Chodos $^1$ and Hisakazu Minakata $^2$}
\medskip
\centerline{$^1$Center for Theoretical Physics, Yale University}
\centerline{217 Prospect Street, New Haven, CT 06511-8167 USA}
\medskip
\centerline{$^2$Department of Physics, Tokyo Metropolitan University}
\centerline{1-1 Minami-Osawa, Hachioji, Tokyo 192-03, Japan}

\bigskip
\bigskip

\centerline{\bf Abstract}
We compute the $1/N$ correction to the location of the previously
found first-order phase transition in the Gross-Neveu model at a
chemical potential $\mu = \mu_c = {1 \over \sqrt{2}} m$, where $m$
is the fermion mass.  We employ an expression for the free energy
$f(\mu)$ given by the thermodynamic Bethe ansatz under the
approximation that the fundamental fermions dominate the ground
state, and combine it with the effective potential evaluated at
zero chemical potential. Our result is
        $\mu_c = {m \over \sqrt{2}} [1 - {0.47 \over N}]$.

\Date{October 15, 1996}

\newsec{Introduction}

In a previous paper \ref\CM{A. Chodos and H. Minakata, Phys. Lett.  
{\bf A191} (1994) 39.}, we studied the Gross-Neveu model 
\ref\GN{D.J. Gross and A. Neveu, Phys. Rev. {\bf D10} (1974) 3235.}  
in the presence of a chemical potential $\mu$.  We computed the effective 
potential analytically to leading order in $1/N$, and we found a 
first-order phase transition at

$$
\mu = \mu_c = {1 \over \sqrt{2}} m  . \eqno(1)
$$

\noindent
where $m$ is the fermion mass.  For $\mu > \mu_c$ the discrete chiral  
symmetry is restored and the fermion becomes massless. The earlier 
references on the phase transition of the Gross-Neveu model may be 
found in ref. \ref\phaset{R. F. Dashen, S.-K. Ma, and R. Rajaraman,  
Phys. Rev. {\bf D11} (1975) 1499;
U. Wolff, Phys. Lett. {\bf B157} (1985) 303.}.

It is known \ref\TLM{H. Takayama, Y. R. Lin-liu, and K. Maki, Phys.  
Rev. {\bf B21} (1980) 2388.} \ref\CB{D. K. Campbell and A. R. Bishop,  
Nucl. Phys. {\bf B200} (1982) 297.} that the $N=2$ Gross-Neveu model  
provides an approximate description of trans-polyacetylene 
\ref\SSH{W. P. Su,  J. R. Schrieffer, and A. J. Heeger, Phys. Rev.
{\bf B22} (1980) 2099;
A. J. Heeger, S. Kivelson, J. R. Schrieffer, and W. P. Su, Rev. Mod.  
Phys.{\bf 60} (1988) 781.}, and we used our results to explain an 
observed phase transition in 
polyacetylene as a function of doping concentration 
(which is related to the chemical potential) \ref\CCMH{J. Chen, T.-C.  
Chung, F. M. Moraes, and A. J. Heeger, Solid State Commun.
{\bf 53} (1985) 757;
F. M. Moraes, J. Chen, T.-C. Chung, and A. J. Heeger, Synth. Met.
{\bf 11} (1985) 271.}.   
We found agreement with the order, the nature, and the location of  
the phase transition, and we concluded that the Gross-Neveu model did  
indeed capture some of the essential physics of trans-polyacetylene.

One aspect of this work that requires further amplification, however,  
is the fact that polyacetylene corresponds to the $N=2$ Gross-Neveu  
model, whereas our results were derived only to leading order in  
$1/N$.  This raises the possibility that the next leading term in the  
$1/N$ expansion might provide a quantitatively significant  
correction, which would then render the excellent agreement between  
theory and experiment found in leading order merely fortuitous.

In this paper we address the question of the order $1/N$ correction  
to our previous computation.  The straightforward way to do this  
would be to evaluate the Feynman graphs that contribute to order  
$1/N$, incorporating, as we did in leading order, the effect of $\mu$  
into a modification of the fermion propagator.  However, the  
evaluation of the $1/N$ correction to the effective potential even  
when $\mu=0$ is a rather involved affair, and the extension to  
non-zero $\mu$ appears to be prohibitively complicated.  Instead, we  
shall evolve a technique based on the thermodynamic Bethe ansatz  
(TBA) \ref\YY{C. N. Yang and C. P. Yang, J. Math. Phys. {\bf 10}  
(1969) 1115.}
\ref\Zamo{Al. B. Zamolodchikov, Nucl. Phys. {\bf B342} (1990) 695.}.

In the next section, we shall explain this technique and illustrate  
it by rederiving the results of our previous paper.  Then in sections  
3 and 4, we shall extend the computation to next order in $1/N$; for  
simplicity, we shall find it convenient to make a further  
approximation in which we restrict the Gross-Neveu S-matrix to a  
single component.  This will be explained more fully below.  In  
section 5 we shall summarize our results and draw some conclusions.

\newsec{The TBA in Leading Order}

The TBA has been much used recently to discuss the properties of  
exactly integrable models \Zamo, \ref\KM{T. R. Klassen and E. Melzer,  
Nucl. Phys. {\bf B338} (1990) 485;
{\bf B350} (1991) 635.}. 
In this paper, somewhat unconventionally,  
we shall actually use it to determine the thermodynamic properties of  
a particular system. The TBA should be very useful technique in 
discussing thermodynamics of the Gross-Neveu model because all the 
S-matrix elements are known \ref\ZZ{A. B. Zamolodchikov and Al. B.  
Zamolodchikov, Ann. Phys. (NY) {\bf 120}
(1979) 253.}
\ref\SW{R Shankar and E. Witten, Nucl. Phys. {\bf B141} (1978) 349.}
\ref\KT{M. Karowski and H. J. Thun, Nucl. Phys. {\bf B190} [FS3]  
(1981) 61.}.

The basic quantity of interest is the "dressed" excitation energy  
$\tilde{\epsilon}(\theta)$ that is required to promote a single  
particle from the Dirac sea to an energy $m cosh\theta$;  
$\tilde{\epsilon}(\theta)$ includes the effects of the rearrangement  
of the vacuum after the particle has been excited.  At finite  
temperature, the equation for $\tilde{\epsilon}(\theta)$ is  
non-linear, but at zero temperature (which is the case of interest to  
us) it reduces to the linear equation \ref\FNW{P. Forgacs, F.  
Niedermayer, and P. Weisz, Nucl. Phys. {\bf 367} (1991) 123.}

$$
\tilde{\epsilon}(\theta) = \mu - m cosh\theta + \int_{-B}^{B}  
d\theta^{\prime} K(\theta - \theta^{\prime})  
\tilde{\epsilon}(\theta^{\prime})  . \eqno(2)
$$

\noindent
Here the kernel $K$ is the logarithmic derivative of the S-matrix:

$$
K(\theta - \theta^{\prime}) = ({1 \over 2\pi i}) {d ~ln S(\theta -  
\theta^{\prime}) \over d\theta} \eqno(3)
$$

\noindent
and $B$ is determined by the condition

$$
\tilde{\epsilon}(\pm B) = 0  .  \eqno(4)
$$

\noindent
Effectively this condition determines $B$ as a function of the  
chemical potential $\mu$.

Once $\tilde{\epsilon}$ is known, the free energy $f$ of the system  
can be determined as a function of $\mu$ up to an additive constant:

$$
f(\mu) - f(0) = - ({m \over 2\pi}) \int_{-B}^{B} d\theta  
\tilde{\epsilon}(\theta) cosh \theta   . \eqno(5)
$$

\noindent
As shown in ref.\ZZ, the expansion of $K$ in powers of $1/N$  
begins in order $1/N$, so to leading order one ignores the integral  
on the right-hand side of eq.  (2), and one has immediately

$$
\tilde{\epsilon}(\theta) = \mu - m cosh\theta   \eqno(6)
$$

\noindent
and hence

$$
coshB = \mu/m   .   \eqno(7)
$$

\noindent
This is possible only if $\mu \geq m$; if $\mu < m$, no solution for  
$\tilde{\epsilon}$ exists.

\smallskip
>From eq. (5) we find

$$
\eqalign{
f(\mu) - f(0) =& \theta(\mu^2 - m^2) ({m^2 \over 2\pi}) (B - {1 \over  
2} sinh 2B) \cr
=& \theta(\mu^2 - m^2) ({m^2 \over 2\pi}) [ln (\mu/m +  
\sqrt{\mu^2/m^2 - 1}) - \mu/m \sqrt{\mu^2/m^2 - 1}]  .  \cr  
}
\eqno(8) $$

\noindent
On dimensional grounds, $f(0) = - bm^2$ where $b$ is a dimensionless  
constant, i.e.

$$
f(\mu) = - bm^2 + \theta(\mu^2 - m^2) {m^2 \over 2\pi} [ln (\mu/m +  
\sqrt{\mu^2/m^2 - 1}) - \mu/m \sqrt{\mu^2/m^2 - 1}]  .   \eqno(9)
$$

\noindent
At first sight, there appears to be no way that this expression can  
be used to provide evidence for a phase transition at $\mu =  
m/\sqrt{2}$, since $f(\mu)$ is absolutely flat for $\mu < m$.  We  
observe, however, that if a massless phase exists, then its free  
energy $f_0(\mu)$ is given by eq. (9) with m=0:

$$
f_0(\mu) = - \mu^2/2\pi  .   \eqno(10)
$$

\noindent
Provided that $0 < b < 1/2\pi$, the curves $f(\mu)$ and $f_0(\mu)$  
will intersect at

$$
\mu^2/m^2 = 2\pi b ,  \eqno(11)
$$

\noindent
with $f_0(\mu) > f(\mu)$ for $(\mu/m) < 2\pi b$.  One therefore  
predicts a first-order phase transition from the massive phase to the  
massless one at $\mu/m = \sqrt{2\pi b}$.

The one piece of information the TBA does not give us is the value of  
$b$.  For this, we need the effective potential computation, but only  
at $\mu = 0$.  Thus the TBA plus the effective potential at $\mu = 0$  
allows us to determine the phase structure of the theory for $\mu >  
0$.

To make contact with previous work, we note that the free energy as  
given above corresponds to the value of the effective potential $V$  
at its minimum $\sigma_{min}$, and the mass $m$ is equal to $g  
\sigma_{min}$ where $g$ is the $GN$ coupling constant.  We can verify  
this explicitly by appealing to eqn. (20) of ref. \CM, 
from which we obtain

$$
\eqalign{
V_{eff}(\sigma) = {\sigma^2 \over 2N} &+ {\sigma^2 \over 4N} [\theta  
(\sigma^2 - \gamma^2) (ln (\sigma^2/\sigma_{0}^{2}) - 3) +  
\theta(\gamma^2 - \sigma^2) (2 ln {\gamma + \sqrt{\gamma^2 -  
\sigma^2} \over \sigma_0} - 3)] \cr
&- {\gamma \over 2N} \sqrt{\gamma^2 - \sigma^2} \theta(\gamma^2 -  
\sigma^2)  ~. \cr
}
\eqno(12) $$

\noindent
This expression comes equipped with the following comments:
\noindent\smallskip
(i) We have divided $V_{eff}$ of ref. [1] by $N$.  This ensures a  
finite limit as $N \rightarrow \infty$ (with $g^2N$ and $m$ fixed)  
and also is the correct normalization to correspond to $f(\mu)$, as  
we shall see explicitly below;
\noindent\smallskip
(ii) We have omitted the last term of eq. (20) of ref. [1], since it  
is  
precisely cancelled by another contribution that was left out of eq.  
(20), called $\tilde{V}$ in eq. (14) of ref. \CM. Thus 
$V_{eff}$ above includes all the relevant contributions;
\noindent\smallskip
(iii) The parameter $\gamma$ above is $\mu\sqrt{N/\pi}$.
\noindent\smallskip
(iv) We have set the coupling $\lambda = g^2N$ that appears in eq.  
(20) of ref. \CM equal to $\pi$.  This is permissible 
because physics depends only on

$$
\sigma_{min} = \sigma_0 e^{(1 - \pi / \lambda)} \eqno(13)
$$

\noindent
where $\sigma_0$ is an arbitrary renormalization point.  The choice  
$\lambda = \pi$ is convenient because then $\sigma_{min} = \sigma_0$.   
We see that

$$
V_{eff}(\sigma_0) = - {\sigma_0^2 \over 4N} + \theta(\gamma^2 -  
\sigma_0^2) [{\sigma_0^2 \over 2N} ln ({\gamma + \sqrt{\gamma^2 -  
\sigma_0^2} \over \sigma_0}) - {\gamma\sqrt{\gamma^2 - \sigma_0^2}  
\over 2N}] ~. \eqno(14)
$$

\smallskip\noindent
Putting $m^2 = g^2 \sigma_0^2 = {\pi \sigma_0^2 \over N}$, and $\mu^2  
= {\pi\gamma^2 \over N}$ we have

$$
V_{eff} = - {m^2 \over 4\pi} + \theta(\mu^2 - m^2) {m^2 \over 2\pi}  
[ln {\mu + \sqrt{\mu^2 - m^2} \over m} - {\mu \over m} \sqrt{{\mu^2  
\over m^2} - 1}] \eqno(15)
$$

\noindent
which tallies exactly with the expression for $f(\mu)$, as given by  
the TBA, provided $b = 1/4\pi$ (and hence $\mu_c = m/\sqrt{2})$.

Of course, as pointed out above, we did not need the full $V_{eff}$,  
including the effects of $\mu$, to determine $b$.  It would have  
sufficed to know the original $V_{eff}(\sigma)$, computed by Gross  
and Neveu \GN:

$$
V_{GN}(\sigma) = {\sigma^2 \over 2N} + {\sigma^2 \over 4N} [ln  
(\sigma^2/\sigma_0^2) - 3] \eqno(16)
$$
\medskip\noindent
from which $V_{GN}(\sigma = \sigma_0 = \sqrt{{N \over \pi}} m) = -  
{m^2 \over4\pi}$ follows immediately as well.

\vfill\eject

\newsec{Next to Leading Order}

In next order, we of course do not have the explicit form of  
$V_{eff}$ as a function of  $\sigma$ and $\mu$.  For $\mu = 0$, the  
$1/N$ correction was computed long ago by Schonfeld \ref\Sch{J.  
Schonfeld, Nucl. Phys. {\bf B95} (1975) 148.} 
and by Root \ref\Roo{R. G. Root, Phys. Rev. {\bf D11} (1975) 831.}.   
Their results will allow us to extract the $1/N$ correction to the  
value $b = + {1 \over 4\pi}$ found in the previous section.  The  
other ingredient we shall need is the form of $f(\mu)$ obtained from  
the TBA, especially the massless limit $f_0(\mu)$.  We shall consider  
each of these in turn.

For our purposes, it is convenient to follow the formulation of the  
$1/N$ correction $\Delta V$ to the Gross-Neveu model given by 
Schonfeld.  By  
summing Feynman diagrams, he obtains an unrenormalized expression.

$$
\Delta V^{un} = \int_0^{\infty} dx \left\{ ln [{1 \over 2}A + ln  
(\sigma/\sigma_0)] - ln~ ln ({x \over g^2\sigma_0^2}) \right\}  
\eqno(17)
$$

\noindent
where we have set $\lambda = \pi$ in Schonfeld's expression, and  
where $A = Z ~ln ({Z + 1 \over Z - 1})$, and $Z = [1 + {4 g^2\sigma^2  
\over x}]^{1/2}$.  $\Delta V^{un}$ is divergent, and it also has an  
imaginary part.  As Schonfeld explains, the correct physical  
prescription, at least in the range of $\sigma$ we are interested in  
$(\sigma \simeq \sigma_0)$ is to take the real part.   The  
divergences are removed by a renormalization prescription which  
amounts to the conditions

$$
Re \Delta V \mid_{\sigma = \sigma_{0}} = {\partial^2 \over  
\partial\sigma^2} (Re \Delta V)\mid_{\sigma = \sigma_{0}} = 0  ~.  
\eqno(18)
$$

\noindent
Now in leading order, $\sigma_{min} = \sigma_0$ (since $\lambda =  
\pi$), and so we must have

$$
\sigma_{min} = \sigma_0 + {1 \over N} \sigma_1, ~~{\rm with} ~~V =  
V_{GN} + Re\Delta V
\eqno(19)$$

\noindent
and

$$
V(\sigma_{min}) = V_{GN}(\sigma_0 + {1 \over N} \sigma_1) + Re\Delta  
V(\sigma_0 + {1 \over N} \sigma_1)  ~. \eqno(20)
$$

\noindent
But $V^{\prime}_{GN}(\sigma_0) = 0$, and $\Delta V$ is already order  
$1/N$, and so to the order we are working,

$$
\eqalign{
V(\sigma_{min}) =& V_{GN} (\sigma_0) + Re \Delta V(\sigma_0) \cr
=& V_{GN}(\sigma_0) \quad\quad\quad\quad\quad{\rm because ~of ~eq.~  
(18)}  \cr
=& - {\sigma_0^2 \over 4N} \cr
=& - {1 \over 4N} (\sigma_{min} - {1 \over N} \sigma_1)^2 = - {1  
\over 4N} (\sigma_{min}^{2} - {2 \over N} \sigma_{min}\sigma_1)  ~.  
\cr
}
\eqno(21)
$$

\noindent
The quantity $\sigma_1$ is determined from the condition

$$
V^{\prime}(\sigma_0 + {1 \over N} \sigma_1) = 0   \eqno(22)
$$

\noindent
and is given by

$$
{\sigma_1 \over N} = - Re \Delta V^{\prime}(\sigma_0) \eqno(23)
$$

\noindent
where use has been made of the renormalization condition  
$V^{\prime\prime}(\sigma_0) = 1$.

It can easily be seen from Schonfeld's paper that in terms of the  
unrenormalized expression $\Delta V^{un}$ defined above, 

$$
Re \Delta V^{\prime}(\sigma_0) = Re \Delta V^{\prime un}(\sigma_0) -  
{4 \over 3\sigma_0} Re \Delta V^{un}(\sigma_0) - {1 \over 3} \sigma_0  
\Delta V^{\prime\prime un}(\sigma_0) \eqno(24)
$$

\noindent
where the last two terms on the $rhs$ are the counterterms necessary  
to implement the conditions (18) and thereby render the expression  
finite.

The remaining task is the purely numerical one of using the integral  
expression (17) to evaluate the right-hand side of eq. (24).  One  
finds

$$
{\sigma_1 \over N} = ({1.06 \over 3N}) \sigma_0  \eqno(25)
$$

\noindent
and hence

$$
V(\sigma_{min}) = - {1 \over 4N} \sigma_{min}^{2} + {1 \over 2N^2}  
({1.06 \over 3}) \sigma_{min}^{2}  \eqno(26)
$$
\smallskip\noindent
where in the second term we have used $\sigma_{min} = \sigma_0 +  
\vartheta(1/N)$.
\medskip
Hence $V_{min} =  - b m^2$ where $b = {1 \over 4\pi}[1 - {2.12 \over  
3N}]$.

\vfill\eject

\newsec{The TBA in Next to Leading Order}

Let us formulate the TBA in the framework of $1/N$ expansion.  While  
we actually deal only with the next to leading order in this paper,  
we present the formulas in an organized way so that the structure of  
the higher orders in $1/N$ becomes transparent.  We expand various  
quantities introduced in section 2 in powers of $1/N$:

$$
\eqalign{
\tilde{\epsilon}(\theta) &= \tilde{\epsilon}_0(\theta) + {1 \over N}  
\tilde{\epsilon}_1(\theta) + \cdot\cdot\cdot \cr
f(\mu) - f(0) &= g_0(\mu) + {1 \over N} g_1(\mu) + \cdot\cdot\cdot  
\cr
K(\theta) &= {1 \over N} K_1(\theta) + {1 \over N^2} K_2(\theta) +  
\cdot\cdot\cdot \cr
}
\eqno(27) $$
\smallskip\noindent
One might think that the boundary of the integral $B$ in (2)  
determined by (4) receives a $1/N$ correction and that it is  
therefore incumbent on us to expand $B$ in powers of $1/N$. 
However, by examining a model in which the TBA equation can be 
solved exactly, we learn that it is more appropriate to keep $B$ 
fixed and change $\mu$ as $N$ is varied. We will treat this in 
the Appendix. Thus we expand the chemical potential

$$
\mu(B) = \mu_0(B) + {1 \over N} \mu_1(B) + \cdot\cdot\cdot  ~~~.   
\eqno(28)
$$
\smallskip\noindent
$\mu_0$ is determined by the leading order equation

$$
\tilde{\epsilon}_0(B) = 0 ~~~.   \eqno(29)
$$
\smallskip\noindent
For clarity we recapitulate the leading order result obtained in  
section 2 in the present notation:

$$
\eqalign{
\tilde{\epsilon}_0 &= \mu_0 - m cosh \theta \cr
g_0(\mu) &= \theta(\mu_{0}^{2} - m^2) {m^2 \over 2\pi} (B - {1 \over  
2} sinh 2B) ~~~.\cr
}
\eqno(30) $$

\bigskip
\noindent
Following Forgacs, Niedermayer and Weisz \FNW, we make the 
ansatz that the  
ground state of the Gross-Neveu model at finite density and zero  
temperature is dominated at large-$N$ by the fundamental fermions.  
This is a reasonable assumption because the solitons and their bound  
states are more massive than the fundamental fermions at large $N$.   
This ansatz allows us to ignore all the S-matrix elements but the  
symmetric combination of the fundamental fermion scattering  
amplitudes, as shown in \FNW.  The approximation is in fact 
implicit  in (2); without the ansatz $K$ becomes a matrix which acts 
on a vector $\tilde{\epsilon}(\theta)$ in the particle species basis.

The consistency of the ansatz is checked by Forgacs, et al., in refs.  
\FNW ~and \ref\FNW2{P. Forgacs, F. Niedermayer, and P. Weisz, Nucl.  
Phys. {\bf 367} (1991) 144.}. The perfect matching of the large-$N$ 
and the TBA results in the leading order, as discussed in the  
preceding 
section, lends further support to this viewpoint.

With this ansatz, the kernel $K$ in the TBA equation (3) in leading  
non-vanishing order has the form

$$
K_1 (\theta) = - {d \over d \theta} ({1 \over \theta} - {1 \over sinh  
\theta})  .  \eqno(31)
$$

\bigskip
\noindent
Using (2) it is straightforward to compute the $1/N$ correction of  
the dressed particle energy $\tilde{\epsilon}(\theta)$ and the free  
energy $f(\mu) - f(0)$. They read

$$
\eqalign{
\tilde{\epsilon}_1 = &\mu_1 + m sinh~ \theta [Chi(\theta + B) -  
Chi(\theta - B) - ln\left\{ {sinh(\theta + B) \over sinh(\theta - B)}  
\right\}] \cr
&- m cosh ~\theta[Shi(\theta + B) - Shi(\theta - B) - 2B] \cr
g_1(\mu) = &- \theta(\mu_{0}^{2} - m^2) {m^2 \over \pi} ~[B^2 +  
sinh^2 B - BShi(2B)] \cr
}
\eqno(32) $$
\bigskip
\noindent
where $Shi(x)$ and $Chi(x)$ are the hyperbolic integral functions  
defined by

$$
\eqalign{
Shi(x) &\equiv \int_{0}^{x} dt {sinh ~t \over t} \cr
Chi(x) &\equiv \gamma + lnx + \int_{0}^{x} dt {cosh ~t - 1 \over t}  
~. \cr
}
\eqno(33) $$

\bigskip
\noindent
Here $\gamma$ is Euler's constant.  The result for $g_1(\mu)$ in (32)  
reproduces that of Forgacs, et al. \FNW .

The boundary condition $\tilde{\epsilon}(B) = 0$ with  
$\tilde{\epsilon}_0(B) = 0$, which is already met in the leading  
order, determines $\mu_0$ and $\mu_1$:

$$
\eqalign{
\mu_0 &= m cosh B \cr
\mu_1 &= m cosh B~[Shi(2B) - 2B~] - m sinh B~[Chi(2B) - ln\left\{  
sinh 2B \right\} - \gamma] \cr
}
\eqno(34) $$

\medskip\noindent
To implement the strategy explained in section 2 we need the  
expression for the free energy in the massless limit, which  
corresponds to $B \rightarrow \infty$. Using the asymptotic form of  
the hyperbolic integral functions

$$
\eqalign{
Shi(x) \rightarrow_{x\rightarrow \infty}{} {e^x \over 2x} \cr
Chi(x) \rightarrow_{x\rightarrow \infty}{} {e^x \over 2x} \cr
}
\eqno(35) $$

\medskip\noindent
one can easily show that

$$
\eqalign{
\lim_{m\rightarrow 0}{} &g_1(\mu) = 0 \cr
\lim_{m\rightarrow 0}{} &\mu_1 = \mu_0(\gamma - ln2) ~~~.\cr
}
\eqno(36) $$
\medskip\noindent
The latter implies that

$$
\mu = \mu_0 (1 + {\gamma - ln2 \over N}) ~~~.   \eqno(37) 
$$
\medskip\noindent
Therefore, we obtain

$$
\eqalign{
\lim_{m\rightarrow 0}{} [f(\mu) - f(0)] &= - {\mu_{0}^{2} \over 2\pi}  
\cr
&= - {\mu^{2} \over 2\pi} [1 + {2 \over N} (ln2 - \gamma] \cr
&= - {\mu^{2} \over 2\pi} (1 + {0.232 \over N}) \cr
}
\eqno(38) $$
\medskip\noindent
We deduce by combining (26) and (38), that the phase transition takes  
place at

$$
{\mu^2 \over 2\pi} (1 + {0.232 \over N}) = {m^2 \over 4\pi} (1 -  
{2.12 \over 3N})  \eqno(39) 
$$

\noindent
or

$$
{\mu^2 \over m^2} = {1 \over 2} (1 - {0.94 \over N})  \eqno(40) 
$$
\medskip\noindent
If we estimate (40) at $N=2$ we have

$$
\mu = {m \over \sqrt{2}} (1 - 0.23)  \eqno(41) 
$$
\medskip\noindent
which implies about a 20\% correction to the critical chemical  
potential due to the next to leading order in $1/N$.

\vfill\eject

\newsec{Conclusions}

In our previous paper, in leading order in the $1/N$ expansion, we  
found a first-order phase transition in the Gross-Neveu model as a  
function of chemical potential, occurring at $\mu_c = {1 \over  
\sqrt{2}} m$.

In this paper, making use of the $1/N$ corrections to the $\mu = 0$  
effective potential, and also of the thermodynamic Bethe ansatz, we  
have shown that the above result receives a correction of the form

$$
\mu_c = {1 \over \sqrt{2}} m [1 - {(.47) \over N}] ~~~.   \eqno(42) 
$$
\medskip\noindent
In view of the result, it is reassuring to note that the correction 
we have found acts to lower the value of  $\mu_c$, i.e. to make it 
easier for the system to undergo the phase transition. The validity 
of our picture might begin to run into trouble if $\mu_c$ were to 
become larger and were to approach $m$ in magnitude, because for 
$\mu > m ~~f(\mu)$ decreases rapidly from its constant value of  
$f(0)$, 
and the question of whether $f(\mu)$ and $f_0(\mu)$ intersect at all  
would then become more problematical.

When we evaluate our correction for $N=2$ we find a 20\% effect.   
Since the value of the doping concentration $y$ is, in this model,  
directly proportional to $\mu$, this implies that $y$ changes from  
.064 to perhaps .050, which is in somewhat less good agreement with  
published results, but still well within the experimental limits.

In obtaining the above result, we have made one further  
approximation:  
following Forgacs, et al., we have restricted the S-matrix appearing 
in the TBA equation to a single component, the scattering between 
fundamental fermions. We believe that it is a good approximation at 
large-N because all other excitations but fundamental fermions have 
masses of the order of N. In fact, the result by Forgacs et al. 
obtained under the same approximation indicates a perfect matching 
between the TBA and the perturbative calculations even at large  
values 
of chemical potential. Nonetheless, we should mention that the  
validity 
of the approximation has to be checked by more elaborate computation 
which includes all the solitons and bound states. The obstacle at 
present to such computation is a considerable amount of complexity  
and 
labor.

Also as we noted above, we have chosen what may seem a curious 
strategy in performing the $1/N$ expansion for the TBA. Rather than 
keeping the physical parameter $\mu$ fixed, and developing the  
auxiliary variable $B$ in a series in $1/N$, we have inverted this  
procedure.  Of course, the two must ultimately be equivalent, but we  
have found, in studying a simple exactly solvable example of the TBA  
equation, that the latter is more transparent and more manageable.  
We amplify this point in a brief appendix.

Another aspect we have discussed in this paper is how the  
density phase transition between massive and massless phases 
can be addressed within the framework of the TBA. The TBA itself 
only allows us to evaluate the difference between free energies 
in finite and zero chemical potential theories. We have proposed 
a new machinery of combining the massless limit of  the TBA result  
with 
that of the computation of the effective potential at zero chemical  
potential. We have verified that it works in the leading order in 
the 1/N expansion of the Gross-Neveu model. It remains to be seen if 
it has wider applicability. It would also be a challenging task 
to see whether one can develop a formalism that could reveal this  
type of 
phase transition solely within the framework of the TBA.

\bigbreak\bigskip\noindent{\bf Acknowledgements}

We thank M. Lippert for help in carrying out the numerical work at  
the end of section 3. This work has been partially supported by  
grants from the National 
Science Foundation INT-9315143, and by Grant-in-Aid for Scientific 
Research under International Scientific Research Program; Joint 
Research No. 07044092 and by the DOE under grant number  
DEFG0292ER40704. It was performed as an activity under support 
by Agreement between Tokyo Metropolitan University and Yale  
University 
on Exchange of Scholars and Collaborations, engaged in May 1996. 
One of us (H.M.) expresses gratitude to Center for Theoretical  
Physics, 
Yale University for warm hospitality extended to him during his  
visits. The other (A.C.) is equally grateful for hospitality extended  
to him during his visits to TMU.

\appendix{A}{}

We give a brief description of an exactly solvable model of the TBA.  
The model is defined by the 2-body S-matrix

$$
S_{ab} = exp ({2\pi i \lambda \over m^2} \epsilon_{\mu\nu}p_{a}^{\mu}  
p_{b}^{\nu})  ,   \eqno(A1)
$$

\noindent 
where $\lambda$ is a real constant.  The S-matrix satisfies  
unitarity and "crossing" but does not have polynomial boundedness,  
and hence may not be realized in local quantum field theories.  We  
shall utilize the model as a theoretical laboratory to examine the  
large-N expansion of the TBA.  We nevertheless keep in mind the  
possibility that it may allow a physical interpretation.  Properties  
of models without polynomial boundedness have recently been discussed  
by Khuri \ref\Khuri{N.N. Khuri, hep-ph/9405406, hep-ph/9512386 (to be  
published in proceedings of the "VIth Blois Workshop, Frontiers in  
Strong Interactions," June 1995).}.  With the definition $p_{i}^{\mu}  
= m(cosh\theta_i, sinh \theta_i)$, $i = a, b$, and the introduction  
of the relative rapidity $\theta = \theta_a - \theta_b$, the kernel  
$K$ takes the simple form

$$
K(\theta) = \lambda cosh \theta    .  \eqno(A2)
$$

\noindent
The TBA equation (2) can be readily solved as

$$
\tilde{\epsilon}(\theta) = \mu - \tilde{m} cosh \theta   \eqno(A3)
$$

\noindent
where

$$
\tilde{m} = {m \over 1 + \lambda [{1 \over 2} sinh 2B - B]}   
\eqno(A4)
$$

\noindent
Notice that $B$ is determined by (4), $\mu - \tilde{m} cosh B = 0$,  
and we can obtain $B$ nonperturbatively in this model. The free  
energy of the model can be easily computed as

$$
f(\mu) - f(0) = - \theta(\mu - \tilde{m})({m^2 \over 2\pi}) {1 \over  
\lambda + [{1 \over 2} sinh 2B - B]^{-1}}  . \eqno(A5)
$$

We treat the model perturbatively to gain insight for the large N  
expansion in the TBA.  We expand various quantities in powers of  
$\lambda$:

$$
\eqalign{
\mu =&~ \mu_0 + \lambda\mu_1 + \lambda^2\mu_2 + \cdot\cdot\cdot. \cr
\tilde{\epsilon}(\theta) =&~ \tilde{\epsilon}_0(\theta) +  
\lambda\tilde{\epsilon}_1(\theta) + \lambda^2  
\tilde{\epsilon}_2(\theta) + \cdot\cdot\cdot. \cr
f(\mu) - f(0) =&~ g_0(\mu) + \lambda g_1(\mu) + \lambda^2g_2(\mu) +  
\cdot\cdot\cdot. \cr
}   \eqno(A6)
$$
\smallskip\noindent
But we keep $B$ fixed as in our treatment of the Gross-Neveu model in  
section 4.  Then one can solve the TBA equation order by order in  
perturbation theory to obtain the results:

$$
\eqalign{
\mu_n =&~ \mu_0[B - {1 \over 2} sinh 2B]^n , \cr
\tilde{\epsilon}_n(\theta) = &~ \tilde{\epsilon}_0(\theta) [B - {1  
\over 2} sinh 2B]^n , \cr
}   \eqno(A7)
$$

\noindent
where

$$
\eqalign{
\mu_0 =&~ m cosh B , \cr
 \tilde{\epsilon}_0(\theta) =&~ \mu_0 - m cosh \theta = m(cosh B -  
cosh \theta) . \cr
}   \eqno(A8)
$$

\smallskip\noindent
It can be readily seen that they reproduce the exact solution (A3)  
and  (A4).

On the other hand, the perturbative treatment of the model when $B$  
is expanded as a power series in $\lambda$ is far from transparent  
and does not appear to work, as may be discovered by making the  
attempt.

There is a whole family of models (of which the above is the simplest  
member) for which the TBA can be solved in similar fashion.

\listrefs

\end